\def\be{\begin{eqnarray}}
\def\ee{\end{eqnarray}}
\def\ba{\begin{array}}
\def\ea{\end{array}}
\def\p{\phi}
\def\pa{\partial }
\def\M{{\cal M}}
\def\G{{\cal G}}
\def\B{{\cal B}}
\def\A{{\cal A}}
\def\X{{\cal X}}
\def\H{{\cal H}}
\def\S{{\cal S}}

\def\K{{\cal K}}
\def\L{{\cal L}}
\def\E{{\cal E}} 
\def\R{{\cal R}}
\def\T{{\cal T}}

\def\C{{\cal C}}
\def\D{^{(D)}}

\documentstyle{article}
\begin{document}

\begin{titlepage}

\centering

{\Large \bf Matrix Ernst Potentials and Orthogonal Symmetry\\
for Heterotic String in Three Dimensions
}
\vskip  2truecm
{\large\bf Alfredo Herrera--Aguilar}\\
\medskip
Laboratory of Computing Techniques and Automation\\
Joint Institute for Nuclear Research, Dubna, M.R. 141980 Russia\\
e--mail: alfa@cv.jinr.dubna.su\\
\bigskip
{\rm and}\\
\bigskip
{\large\bf Oleg Kechkin}\\
Department of Electromagnetic Processes and Nuclear Interactions\\
Nuclear Physics Institute, Moscow State University, 
Moscow 119899 Russia \\
e--mail: kechkin@monet.npi.msu.su

\vskip 2truecm

{\large April 1997}

\vskip 2truecm

\begin{abstract}
A new coset matrix for low--energy limit of
heterotic string theory reduced to three dimensions is constructed. The
pair of matrix Ernst potentials uniquely connected with the coset matrix is
derived. The action of the symmetry group on the Ernst potentials is
established.
\end{abstract}

\end{titlepage}
\section{Review of Previous Results}
In one--loop approximation the heterotic string theory leads to the effective
action which describes matter fields coupled to gravity \cite{ms}:
\be
S\D [G\D_{MN},B\D_{MN},\p\D,A\D_{M}]=  
\int d\D x \mid G\D\mid^{\frac{1}{2}}\,\hspace{-6mm}&&e^{-\p\D} 
(R\D + 
\p\D_{;M} \p^{(D);M} -
\nonumber
\\
&&\frac{1}{12}\,H\D_{MNP}\, H^{(D)MNP} - 
\frac{1}{4}\,F^{(D)I}_{MN}\, F^{(D)IMN}),
\ee
where 
\be
&&F^{(D)I}_{MN}=\pa _MA^{(D)I}_N-\pa _NA^{(D)I}_M, 
\nonumber
\\
&&H\D_{MNP}=\pa _MB\D_{NP}-\frac{1}{2}A^{(D)I}_M\,F^{(D)I}_{NP}+   
\mbox{\rm cycl. perms. of M,N,P.}
\nonumber
\ee
Here $G\D_{MN}$ is the $D$-dimensional metric, $B\D_{MN}$ is the 
antisymmetric Kalb--Ramond field, $\p\D$ is the dilaton and $A^{(D)I}_M$ 
denotes a set ($I=1,\,2,\,...,n$) of Abelian vector fields. For the 
self--consistent heterotic string theory $D=10$ and $n=16$ \cite {s}, 
but in this work, following \cite {ms}, we shall leave these parameters 
arbitrary.

The action (1) can be generalized for the case of Yang--Mills gauge fields; 
it can also include mass, Gauss--Bonnet terms, etc. But only the simplest 
variant (1) of the theory possesses remarkable analytical properties 
which are important for our consideration.

In \cite {ms}--\cite {s} it was shown that after the Kaluza--Klein 
compactification of $d=D-3$ dimensions on a torus, the resulting theory is
\be
S^{(3)}[g_{\mu\nu},B_{\mu\nu},\p,A_{\mu},M]= 
\int d^3 x \mid g\mid^{\frac{1}{2}}\, 
[R + \p_{;\mu} \p^{;\mu} - \hspace{-6mm}
&&\frac{1}{12}\,H_{\mu\nu\rho}\, H^{\mu\nu\rho} - 
\nonumber
\\
&&e^{-2\p}F^T_{\mu\nu}M^{-1}F^{\mu\nu}-
\frac{1}{8}\,Tr\left(J^M\right)^2].
\ee
Here the symmetric matrix $M$ has the following structure
\be
M=\left(
\ba{ccc}
G^{-1} & G^{-1}(B+C) & G^{-1}A \cr 
(-B+C)G^{-1} & (G-B+C)G^{-1}(G+B+C) & (G-B+C)G^{-1}A \cr  
A^{T}G^{-1} & A^{T}G^{-1}(G+B+C) & I_n+A^{T}G^{-1}A
\ea
\right)
\ee
with block elements defined by
\be 
&&G=(G_{pq} \equiv G\D_{p+2,q+2}), 
\nonumber
\\
&&B=(B_{pq} \equiv B\D_{p+2,q+2}), 
\nonumber
\\
&&A=(A^I_p \equiv A^{(D)I}_{p+2}), 
\nonumber
\ee
where $C=\frac{1}{2}AA^{T}$ and $p,q=1,2,...,d$. Matrix $M$ satisfies the 
$O(d,d+n)$ group relation
\be
MLM=L,
\ee
where
\be
L=\left(
\ba{ccc}
O & I_d & 0  \cr
I_d & 0 & 0  \cr
0 & 0 & -I_n
\ea
\right);
\ee
thus $M\in O(d,d+n)/O(d)\times O(d+n)$.

The remaining $3$--fields are defined in the following way:
for dilaton and metric fields one has 
\be
&&\p=\p\D-\frac{1}{2}ln\,det\,G,  
\nonumber
\\
&&g_{\mu\nu}=e^{-2\p}\left(G\D_{\mu\nu}-G\D_{p+2,\mu} G\D_{q+2,\nu}G^{pq}
\right).
\nonumber
\ee
Then, the set of Maxwell strengths $F^{(a)}_{\mu\nu}$ ($a=1,2,...,2d+n$) is 
constructed on $A^{(a)}_{\mu}$, where 
\be
&&A^p_{\mu}=\frac{1}{2}G^{pq}G\D_{q+2,\mu}  
\nonumber
\\
&&A^{I+2d}_{\mu}=-\frac{1}{2}A^{(D)I}_{\mu}+A^I_qA^q_{\mu}, 
\nonumber
\\
&&A^{p+d}_{\mu}=\frac{1}{2}B\D_{p+2,\mu}-B_{pq}A^q_{\mu}+
\frac{1}{2}A^I_{p}A^{I+2d}_{\mu}.
\nonumber
\ee      
Finally, the $3$--dimensional axion 
$$
H_{\mu\nu\rho}=\pa_{\mu}B_{\nu\rho}+2A^a_{\mu}L_{ab}F^b_{\nu\rho}+
\mbox{\rm cycl. perms. of $\mu$, $\nu$, $\rho$}
$$ 
depends on the $3$--dimensional Kalb--Ramond field 
$$
B_{\mu\nu}=B\D_{\mu\nu}-4B_{pq}A^p_{\mu}A^q_{\nu}-
2\left(A^p_{\mu}A^{p+d}_{\nu}-A^p_{\nu}A^{p+d}_{\mu}\right).
$$

The dimensionally reduced system (2) admits two simplifications. Namely, in three dimensions, the
Kalb--Ramond field $B_{\mu\nu}$ becomes a non--dynamical variable and can be
omitted \cite {s}. Moreover, the fields $A_\mu^a$ can be dualized on--shell
as follows
\be
e^{-2\p}MLF_{\mu\nu}=\frac{1}{2}E_{\mu\nu\rho}\nabla^{\rho}\psi;
\ee
so, the final system is defined by the quantities $M$, $\p$ and $\psi$. As
it had been established by Sen in \cite {s}, it is possible to introduce the
matrix
\be
\M_S=\left(
\ba{ccc}
M+e^{2\p}\psi\psi^T & -e^{2\p}\psi & ML\psi+\frac{1}{2}\psi(\psi^{T}L\psi) \cr 
-e^{2\p}\psi^T & e^{2\p} &  -\frac{1}{2}e^{2\p}\psi^{T}L\psi  \cr  
\psi^TLM+\frac{1}{2}e^{2\p}\psi^T(\psi^{T}L\psi) & 
-\frac{1}{2}e^{2\p}\psi^{T}L\psi & 
e^{-2\p}+\psi^TLML\psi+\frac{1}{4}e^{2\p}(\psi^{T}L\psi)^2
\ea
\right),
\ee
in terms of which the action of the system adopts the standard chiral form
\be
\ba{l}
S^{(3)}[g_{\mu\nu},\M_S]= 
\int d^3 x \mid g\mid^{\frac{1}{2}}\, 
\left[R - \frac{1}{8}\,Tr\left(J^{\M_S}\right)^2\right],
\ea                                                  
\ee
where $J^{\M_S}=\nabla\M_S\M^{-1}_S$.
This matrix is symmetric $\M_S=\M^T_S$ and satisfies the  
$O(d+1,d+n+1)$--group relation
\be
\M_S\L_S\M_S=\L_S
\ee
with
\be
\L_S=\left(
\ba{ccc}
L & 0 & 0  \cr
0 & 0 & 1  \cr
0 & 1 & 0
\ea
\right),
\ee
so that, $\M_S$ belongs to the coset $O(d+1,d+n+1)/O(d+1)\times O(d+n+1)$.

It is easy to see that the coset $O(d+1,d+n+1)/O(d+1)\times O(d+n+1)$
can be obtained from the coset $O(d,d+n)/O(d)\times O(d+n)$
by the replacement $d\rightarrow d+1$. At the same time, $\M_S$ has a quite 
different structure in comparison with $M$. Making use of these facts, one 
can hope that there is another chiral matrix $\M$ possessing
the same structure that $M$ with block components $\G$, $\B$ and $\A$ of 
$(d+1)\times (d+1)$, $(d+1)\times (d+1)$ and $(d+1)\times n$ dimensions, 
respectively.

In this paper we show that such a matrix can actually be constructed. 
We establish that its block components allow to define two matrices 
(``matrix Ernst potentials") which permit to represent the theory under 
consideration in the Einstein--Maxwell (EM) form. At the end of the paper 
we study how the $O(d+1,d+n+1)$ group of transformations acts on the 
matrix Ernst potentials and establish the relations between its subgroups 
on the base of the discrete strong--weak coupling duality transformations 
(SWCDT) found in \cite {s}.

\section{Matrix Ernst Potentials}

We start from the consideration of the kinetic term of the matrix $M$
\be
\ba{l}
S^{(3)}[M]=-\frac{1}{8}\int d^3 x \mid g\mid^{\frac{1}{2}}\, 
Tr\left(J^{M}\right)^2.
\ea                                                  
\ee
The Euler--Lagrange equation corresponding to (11) is
\be
\nabla J^{M}=0.
\ee
In terms of the block components $G$, $B$ and $A$ it reads
\be
&&\nabla J^{G}-\left(J^B\right)^2 +\nabla A\nabla A^T G^{-1}=0, 
\nonumber
\\
&&\nabla J^{B}-J^GJ^B=0, 
\nonumber
\\
&&\nabla \left(G^{-1}\nabla A\right)-G^{-1}J^B\nabla A=0, 
\ee
where
\be
&&J^G=\nabla G\,G^{-1}, 
\nonumber
\\
&&J^B=\left[\nabla B+
\frac{1}{2}\left(A \nabla A^T-\nabla AA^T\right)\right]G^{-1}.
\ee
Eqs. (13) are the motion equations for the action
\be
S^{(3)}[G,B,A]=-\int d^3 x \mid g\mid^{\frac{1}{2}}\, 
Tr\left\{\frac{1}{4}\left[\left(J^{G}\right)^2-\left(J^{B}\right)^2\right]+
\frac{1}{2} \nabla A^T\,G^{-1}\nabla A\right\},
\ee
which is equivalent to (11) and can be obtained by straightforward but tedious
algabraical calculations. (The coefficient $\frac{1}{4}$ can easily be
established by comparison of Eqs. (11) and (15) in the case when $B=A=0$).

One can introduce the matrix variable
\be
X=G+B-\frac{1}{2}AA^T,
\ee
which was entered for the first time by Maharana and Schwartz in the case 
when $A=0$ \cite {ms}; it defines, together with $A$, the most compact 
constrainless representation of the system:
\be
\ba{l}
S^{(3)}[X,A]=-\int d^3 x \mid g\mid^{\frac{1}{2}}\, 
Tr\left[\frac{1}{4}\left(\nabla X+A\nabla A^T\right)G^{-1}
\left(\nabla X^T+\nabla AA^T\right)G^{-1}+ 
\frac{1}{2}\nabla A^T\,G^{-1}\nabla A\right],
\ea
\ee
where $G=\frac{1}{2}\left(X+X^T-AA^T\right)$. The form of this action is very 
similar to the stationary Einstein--Maxwell one \cite {iw}--\cite {m}. 
Thus, in string gravity the 
matrix $X$ formally plays the role of the gravitational potential $\E$, 
whereas the matrix $A$ corresponds to the electromagnetic potential $\Phi$ 
of EM theory \cite{e}. At the same time, one can notice a direct 
correspondence between the transposition of $X$ and $A$ on the one hand, 
and the complex conjugation of $\E$ and $\Phi$, on the other. This analogy 
will be useful to study the symmetry group of string gravity in the last 
chapter of the paper.

For the complete theory, i.e., for the theory with nontrivial fields $\p$ 
and $\psi$, the chiral current $J^M$ does not preserve and one has the 
equation
\be
\nabla J^M+4e^{-2\p}FF^TM^{-1}=0
\ee
instead of (12). The additional $\p$-- and $\psi$--equations of motion are:
\be
&&\nabla ^2\p+\frac{1}{2}e^{2\p}\nabla \psi^TM^{-1}\nabla\psi=0, 
\nonumber
\\
&&\nabla _{\mu}\left(e^{-2\p}M^{-1}F^{\mu\nu}\right)=0.
\ee
They can be derived from the action
\be
S^{(3)}[M,\p,\psi]=-\int d^3 x \mid g\mid^{\frac{1}{2}}\, 
Tr\left[(\nabla \p)^2-\frac{1}{2}e^{2\p}\nabla \psi^TM^{-1}\nabla\psi+
\frac{1}{8}Tr\left(J^M\right)^2\right]
\ee
by the usual variational procedure.

Our main aim is to represent the action (20) in a form similar to (11).
We suppose that it can be done by the 
$[2(d+1)+n]\times [2(d+1)+n]$ matrix $\M$ defined by the block 
components $\G$, $\B$ and $\A$ in the same way that the 
$[2d+n]\times [2d+n]$ matrix $M$ is defined by $G$, $B$ and $A$:
\be
\M=\left(
\ba{ccc}
\G^{-1} & \G^{-1}(\B+\C) & \G^{-1}\A \cr 
(-\B+\C)\G^{-1} & (\G-\B+\C)\G^{-1}(\G+\B+\C) & (\G-\B+\C)\G^{-1}\A \cr  
\A^{T}\G^{-1} & \A^{T}\G^{-1}(\G+\B+\C) & I_n+\A^{T}\G^{-1}\A
\ea
\right).
\ee
This matrix also is a symmetric one and satisfies the $O(d+1,d+n+1)$--group
relation
\be
\M\L\M=\L,
\ee
where
\be
\L=\left(
\ba{ccc}
0 & I_{d+1} & 0  \cr
I_{d+1} & 0 & 0  \cr
0 & 0 & -I_n
\ea
\right),
\ee
and belongs to the coset $O(d+1,d+n+1)/O(d+1)\times O(d+n+1)$.

This hypothesis means that the action (20) can be expressed in the form
\be
S^{(3)}[\M]=-\frac{1}{8}\int d^3 x \mid g\mid^{\frac{1}{2}}\, 
Tr\left(J^{\M}\right)^2
\ee
with $J^{\M}=\nabla\M\M^{-1}$; in view of (21), one can rewrite it as
\be
S^{(3)}[\G,\B,\A]=-\int d^3 x \mid g\mid^{\frac{1}{2}}\, 
Tr\left\{\frac{1}{4}\left[\left(J^{\G}\right)^2-\left(J^{\B}\right)^2\right]+
\frac{1}{2}\nabla \A^T\,\G^{-1}\nabla \A\right\}.
\ee

In order to establish the explicit form of the matrix $\M$ one can procede as
follows. On the one hand, it is useful to represent the column $\psi$ in the
form
\be
\L_S\psi=\left(
\ba{l}
u \cr
v \cr
s
\ea
\right).
\ee
Then Eq. (25) transforms to
\be
S^{(3)}[G,B,A,\p,u,v,s]=-\int d^3 x \mid g\mid^{\frac{1}{2}} 
\{(\nabla\p)^2+Tr\left[\frac{1}{4}
\left((J^{\G})^2-(J^{\B})^2\right)+ 
\frac{1}{2}\nabla \A^T\,\G^{-1}\nabla \A\right]- 
\frac{1}{2}e^{2\p}(\nabla u+
\nonumber
\ee
\vspace{-5mm}
\be
(B+C)\nabla v+A\nabla s)^TG^{-1}
(\nabla u+
(B+C)\nabla v+A\nabla s)+  
\nabla v^TG\nabla v+\left(\nabla s-A^T\nabla v\right)^T
\left(\nabla s-A^T\nabla v\right)\}.
\ee
On the other hand, the parametrization \footnote{The parametrization of
the matrices $G$ and $B$ is written using the analogy between the theory
under consideration and the theories with symplectic symmetry \cite{ky1}--
\cite {ky2}.}
\be
\G=\left(
\ba{cc}
-f+\tilde v^TG\tilde v & \tilde v^TG \cr 
G\tilde v & G
\ea
\right), \quad 
\B=\left(
\ba{cc}
0 & \tilde w^T \cr 
-\tilde w & B
\ea
\right), \quad 
\A=\left(
\ba{c}
\tilde s^T \cr 
A
\ea
\right), 
\ee
with $\tilde w=\tilde u+B\tilde v$, leads to the following expressions
for the 1--st and 3--rd terms of Eq. (25)
\be
\ba{l}
S^{(3)}[\G]=-\frac{1}{4}\int d^3 x \mid g\mid^{\frac{1}{2}}\, 
Tr\left(J^{\G}\right)^2=-\int d^3 x \mid g\mid^{\frac{1}{2}}\, 
\left\{\frac{1}{4}\left[f^{-2}(\nabla f)^2+Tr\left(J^{\G}\right)^2\right]-
\frac{1}{2}f^{-1}\nabla v^TG\nabla v\right\},
\ea
\ee
\be
S^{(3)}[\A]=-\frac{1}{2}\int d^3 x \mid g\mid^{\frac{1}{2}}\, 
Tr\left(\nabla A^T\G^{-1}\nabla A\right)= 
-\frac{1}{2}\int d^3 x \hspace{-6mm}&&\mid g\mid^{\frac{1}{2}}\, 
[Tr\left(\nabla A^TG^{-1}\nabla A\right)-
\nonumber
\\
&&f^{-1}
\left(\nabla\tilde s-\nabla A^T\tilde v\right)^T
\left(\nabla\tilde s-\nabla A^T\tilde v\right)
].
\ee
One can see that Eq. (29) gives the 1--st, 2--nd and 6--th 
terms of Eq. (27) if
\be
f=e^{-2\p} \quad \mbox{\rm and} \quad \tilde v=v.
\ee                                            
On the other hand Eq. (30) is equivalent to the 4--th and
7--th items of Eq. (25) if
\be
\tilde s=-s+A^Tv.
\ee
The second term of Eq. (25)
\be
S^{(3)}[\B]=\hspace{-6mm}&&\frac{1}{4}\int d^3 x \mid g\mid^{\frac{1}{2}}\, 
Tr\left(J^{\B}\right)^2=\int d^3 x \mid g\mid^{\frac{1}{2}}\, 
\left\{\frac{1}{4}Tr\left(J^{\B}\right)^2+\frac{1}{2}f^{-1}
\times\right.
\nonumber
\\
&&\left.\left[\nabla \left(\tilde u-\frac{1}{2}As\right)+
(B+C)\nabla v+A\nabla s\right]^TG^{-1}
\left[\nabla \left(\tilde u-\frac{1}{2}As\right)+
(B+C)\nabla v+A\nabla s\right]\right\}
\ee
corresponds to the remaining 3--rd and 5--th items of Eq. (27) if
\be
\tilde u=u+\frac{1}{2}As.
\ee

Thus, the block components of the matrix $\M$ are defined by Eqs. 
(28), (31), (32) and (34). Consequently, the matrices $\G$ and $\B$ are
\be
\G=\left(
\ba{cc}
-e^{-2\p}+v^TGv & v^TG \cr 
Gv & G
\ea
\right), \qquad 
\B=\left(
\ba{cc}
0 & \tilde w^T \cr 
-w & B
\ea
\right), 
\ee
where $w=u+Bv+\frac{1}{2}As$. Finally, for the matrix Ernst potentials 
$\X$ and $A$  one has
\be
\ba{c}
\X=\left(
\ba{cc}
-e^{-2\p}+v^TXv-v^TAs-\frac{1}{2}s^Ts & v^TX+u^T+s^TA^T \cr 
Xv-u & X
\ea
\right),     \cr
\A=\left(
\ba{c}
-s^T+v^TA \cr 
A
\ea
\right). 
\ea
\ee
\section{Matrix Ehlers--Harrison Transformations}

In this section we establish the action of the symmetry group $O(d+1,d+n+1)$
on the matrix Ernst potentials $\X$ and $\A$. It is evident that the action
\be
S^{(3)}[g_{\mu\nu},\X,\A]=-\int d^3 x \mid g\mid^{\frac{1}{2}}\, 
\left\{R-Tr\left[\frac{1}{4}\left(\nabla \X+\A\nabla \A^T\right)\G^{-1}
\left(\nabla \X^T+\nabla \A\A^T\right)\G^{-1}+ 
\frac{1}{2}\nabla \A^T\,\G^{-1}\nabla \A\right]\right\},
\nonumber
\ee
\vspace{-6mm}
\be
\ee
where $\G=\frac{1}{2}\left(\X+\X^T-\A\A^T\right)$, is invariant under the 
``rotation" 
\be
&&\A=\A_0\H,
\nonumber
\\
&&\X=\X_0,
\ee
where $\H\H^T=I_n$; this map generalizes the duality rotation of 
the electromagnetic sector in the stationary EM theory 
\cite {k}. One can also see that the ``scaling" 
\be
&&\A=\S^T\A_0,
\nonumber
\\
&&\X=\S^T\X_0\S,
\ee
where $det\S\ne 0$, corresponds to the scale transformation of  
EM system. The gauge transformation of the potential $\A$ reads
\be
&&\A=\A_0,
\nonumber
\\
&&\X=\X_0+\R_1
\ee
with $\R^T_1=-\R_1$, whereas for the gauge shift of the potential 
$\X$ one obtains
\be
&&\A=\A_0+\T_1,
\nonumber
\\
&&\X=\X_0-\T_1\A^T_0-\frac{1}{2}\T_1\T^T_1.
\ee
These transformations are the matrix analogues of the shifts of the 
rotational and electromagnetic variables of the stationary EM theory. 

In order to find nontrivial transformations one can use SWCDT \cite {s}.
This symmetry transformation 
\be
\M\rightarrow \M^{-1}
\ee
can be expressed in terms of the matrices $\X$ and $\A$ as follows
\be
&&\A\rightarrow -\left(\X+\A\A^T\right)^{-1}\A, 
\nonumber
\\
&&\X\rightarrow \left(\X+\A\A^T\right)^{-1}\X^T\left(\X^T+\A\A^T\right)^{-1}.
\ee

Using this map it is possible to obtain new transformations from the known
ones (38)--(41). However, the scaling matrix subgroups remain invariant 
($\H\rightarrow H$ and $\S\rightarrow (\S^T)^{-1}$) under (43). It turns 
out that the shift subgroups give rise to the actually non--linear 
transformations
\be
&&\A=\left[1+\left(\X_0+\A_0\A^T_0\right)\R_2\right]^{-1}\A_0, 
\nonumber
\\
&&\left(\X+\A\A^T\right)^{-1}=\left(\X_0+\A_0\A^T_0\right)^{-1}+\R_2,
\ee
where $\R^T_2=-\R_2$, and
\be
&&\A=\left[1+\A_0\T^T_2+\frac{1}{2}\left(\X_0+\A_0\A^T_0\right)
\T_2\T^T_2\right]^{-1}\left[\A_0+\left(\X_0+\A_0\A^T_0\right)T_2\right], 
\nonumber
\\
&&\X+\A\A^T=\left[1+\A_0\T^T_2+\frac{1}{2}
\left(\X_0+\A_0\A^T_0\right)\T_2\T^T_2\right]^{-1}
\left(\X_0+\A_0\A^T_0\right).
\ee
Formula (44) generalizes the Ehlers transformation \cite{eh} for the 
string system, whereas Eq. (45) provides the matrix analogue of the 
Harrison (``charging") transformation \cite {k}.

At the end of the paper we would like to remark that the relations 
(38)--(41) and (44)-(45) form the full set of transformations of 
the $O(d+1,d+n+1)$ group. Actually, the general  $O(d+1,d+n+1)$ 
matrix $\K$, which defines the authomorphism $\M \rightarrow \K^T\M\K$, 
can be represented in the following form
\be
\K=\K_{\T_2}\K_{\R_2}\K_\S\K_\H\K_{R_1}\K_{\T_1},
\ee
where
\be
\ba{cc}
\K_{\T_2}=\left(
\ba{ccc}
I_{d+1} & 0 & 0 \cr 
K_{\T_2} & I_{d+1} & \T_2\cr
\T^T_2 & 0 & I_n
\ea
\right), & \vspace{6mm}
\K_{\R_2}=\left(
\ba{ccc}
I_{d+1} & 0 & 0 \cr 
\R_2 & I_{d+1} & 0 \cr
0 & 0 & I_n
\ea
\right), \cr \vspace{6mm}
\K_{\S}=\left(
\ba{ccc}
(\S^T)^{-1} & 0 & 0 \cr 
0 & \S & 0 \cr
0 & 0 & I_n
\ea
\right),  &   
\K_{\H}=\left(
\ba{ccc}
I_{d+1} & 0 & 0 \cr 
0 & I_{d+1} & \cr
0 & 0 & \H
\ea
\right), \cr 
\K_{\R_1}=\left(
\ba{ccc}
I_{d+1} & \R_1 & 0 \cr 
0 & I_{d+1} & 0 \cr
0 & 0 & I_n
\ea
\right), &  
\K_{\T_1}=\left(
\ba{ccc}
I_{d+1} & K_{\T_1} & T_1 \cr 
0 & I_{d+1} & 0 \cr
0 & \T^T_1 & I_n
\ea
\right). 
\ea
\ee
Here $K_{\T_2}=\frac{1}{2}\T_2\T^T_2$ and $K_{\T_1}=\frac{1}{2}\T_1\T^T_1$;
moreover, $\left[\K_{\T_2},\K_{\R_2}\right]=\left[\K_{\R_1},\K_{\T_1}\right]=
\left[\K_{\S},\K_{\H}\right]=0$, and under the map (42) one has
\be
&&\K_{\T_1}\rightarrow \K_{\T_2}, 
\nonumber
\\
&&\K_{\H}\rightarrow \K_{\H}, 
\nonumber
\\
&&\K_{\S}\rightarrow \K_{(\S^T)^{-1}}, 
\nonumber
\\
&&\K_{\R_1}\rightarrow \K_{\R_2}, 
\ee
where $\R_1\rightarrow \R_2$ and $\T_1 \rightarrow -\T_2$. Thus, the 
complete $O(d+1,d+n+1)$ group consists of six subgroups defined 
by the matrices $\H$, $\S$; $\T_1$, $\T_2$; $\R_1$, $\R_2$. This subgroups 
are the same ones that we have considered above (see Eqs. (38)--(41) and 
(44)--(45)). 
\section{Conclusion and Discussion}

In this paper we study the $O(d+1, d+n+1)$--symmetric low--energy limit 
of heterotic string theory reduced to three dimensions. It is shown 
that such a theory can be represented in terms of the $(d+1)\times(d+1)$ 
matrix $\X$ and $(d+1)\times n$ matrix $\A$. These matrices appear to be 
the analogues of the gravitational and electromagnetic potentials ($\E$ 
and $\Phi$, respectively) of the stationary EM theory. The matrices 
$\G=\frac{1}{2}\left(\X+\X^T-\A\A^T\right)$, 
$\B=\frac{1}{2}\left(\X-\X^T\right)$ and $\A$ define the chiral matrix 
$\M\in O(d+1,d+n+1)/O(d+1)\times O(d+n+1)$ of the theory in the same way 
that matrices $G$, $B$ and $A$ (constructed on the extra components of 
the metric, Kalb--Ramond and electromagnetic fields, respectively) define 
the coset matrix $M\in O(d,d+n)/O(d)\times O(d+n)$.

It is established that the $O(d+1,d+n+1)$ symmetry group can be decomposed
into six subgroups using the strong--weak coupling duality transformation. 
It turns out that two subgroups (the rescaling of the potentials $\X$ and 
$\A$) are invariant under SWCDT. At the same time, the remaining 
transformations combine into two pairs which map one into another under 
SWCDT. We show that the gauge shift of $\X$ maps into the matrix Ehlers 
tranformation, whereas the shift of $\A$ maps into the matrix Harrison one. 

All subgroups of transformations are written in quasi--Einstein--Maxwell 
form. This fact remarks the analogy between the string gravity system with
orthogonal symmetry, on the one hand, and the EM theory, on the other, in 
the $3$--dimensional case.
\section*{Acknowledgments}

We wuold like to thank our colleagues of DEPNI (NPI) and JINR for an 
encouraging relation to our work, as well as to ICTP for the hospitality 
and facilities provided during our stay at Trieste, where the final version 
of this paper was performed. One of the authors (A.H.) was supported in 
part by CONACYT and SEP.                            

\end{document}